 \UseRawInputEncoding
\documentclass[%
reprint,
superscriptaddress,
floatfix,
 amsmath,amssymb,amsfonts,
prb,
]{revtex4-2}

\usepackage{wasysym}
\usepackage{graphics,graphicx,epsfig}
\usepackage{amssymb,color,ulem}
\usepackage{epsf,epstopdf,wrapfig}
\usepackage{mciteplus}
\usepackage{graphicx,color}
\usepackage{amsmath,amssymb}
\usepackage{dcolumn}   
\usepackage{bm}   
\usepackage{multirow}
\usepackage{dsfont}
\usepackage{lipsum,afterpage}
\usepackage[colorlinks=true,citecolor=blue]{hyperref}
\hypersetup{colorlinks=true,citecolor=blue,linkcolor=blue,urlcolor=blue}
\usepackage{soul}
\usepackage{xcolor}
\usepackage{enumitem}
\usepackage{relsize}
\usepackage{cancel}
\def\be{\begin{equation}}
\def\ee{\end{equation}}
\def\bea{\begin{eqnarray}}
\def\eea{\end{eqnarray}}

\def\ra{\rangle}
\def\la{\langle}

\newcommand{\aref}[1]{\hyperref[#1]{Appendix~\ref{#1}}}

\begin{document}

\title{Role of activity and dissipation in achieving precise beating in cilia: Insights from the rower model} 

\author{Subhajit Gupta}
\affiliation{Department of Physics, SRM University - AP, Amaravati, Andhra Pradesh 522240}

\author{Debasish Chaudhuri}
\email[For correspondence:~]{debc@iopb.res.in}
\affiliation{Institute of Physics, Sachivalaya Marg, Bhubaneswar-751005, Odisha, India}
\affiliation{Homi Bhabha National Institute, Anushakti Nagar, Mumbai 400094, India}

\author{Supravat Dey}
\email[For correspondence:~]{ supravat.d@srmap.edu.in }
\affiliation{Department of Physics, SRM University - AP, Amaravati, Andhra Pradesh 522240}

\begin{abstract}
Cilia and flagella are micron-sized filaments that actively beat with remarkable precision in a viscous medium, driving microorganism movement and efficient flow. We study the rower model to uncover how cilia activity and dissipation enable this precise motion. 
In this model, cilia motion is represented by a micro-bead’s Brownian movement between two distant harmonic potentials. At specific locations, energy pumps trigger potential switches, capturing cilia activity and generating oscillations. 
We quantify precision of oscillation using a quality factor, identifying its scaling with activity and oscillation amplitude, finding precision maximization at an optimal amplitude. The data collapse is not accurate for noisy oscillations. An exact analytic expression for the precision quality factor, based on first passage time fluctuations, and derived in the small noise approximation, explains its optimality and scaling. Energy budget analysis shows the quality factor's consistency with the thermodynamic uncertainty relation. 
Finally, we demonstrate that asymmetric beating reduces oscillation precision compared to the symmetric model: although the optimal amplitude remains unchanged, the overall scaling of the quality factor depends explicitly on the asymmetry parameter.
 
%
\end{abstract}

\maketitle

\section{Introduction}
Motile cilia and flagella, hairlike appendages found on many unicellular organisms and epithelial tissues, actively beat to drive movement \cite{Graybook, Sleigh2016, Brennen1977, satir2007overview}. Their structure consists of nine outer microtubule doublets and two central doublets \cite{Porter2000}, with sliding movement powered by dynein motors \cite{julicher2007molecular, Gilpin2020}. This internal force causes microtubules to slide relative to each other \cite{Porter2000, mitchison2010cilia, julicher2016curvature, Bruot2016, king2018fifty, vilfan2019flagella}, generating complex oscillations in the surrounding fluid medium. These movements, often synchronized into metachronal waves, are essential for processes like mucociliary clearance \cite{SmithRes08, Sleigh2016, Ramirez2020} and the locomotion of microorganisms such as Paramecium, Opalina, and Volvox \cite{Tamm_Jcb1975, Machemer_jeb1972, Brumley2012}.

Cilia and flagella exhibit both coordinated movement in large groups and remarkable precision in their individual oscillations, despite fluctuations arising from the intrinsic stochastic activity of molecular motors and external noises from ambient fluid dynamics. The precision of these oscillations is quantified using a quality factor $Q$, a dimensionless ratio of coherence time to oscillation period~\cite{remlein2022coherence, barato2017coherence, bagheri2014effects, Roldan_njp2021}. 
Values of $Q$ typically range from $\sim 30-100$ \cite{placcais2009spontaneous, martin2001comparison, goldstein2009noise, goldstein2011emergence, Ma2014, maggi2023thermodynamic, sharma2024active}, indicating high precision.  In {\it Chlamydomonas reinhardtii}, flagellar precision increases with the number of motor proteins. Experimental findings are summarized in a recent study \cite{sharma2024active}.

A multitude of models, reflecting varying degrees of complexity in the active forces at play, have been proposed to explore the genesis of sustained \cite{yang2008integrative, Guirao2007, gueron1998computation, BratoPRFluid2019, oriola2017nonlinear} and synchronized oscillations \cite{GolestanianRev11, Bruot2016, Gilpin2020, han2018spontaneous, ding2014mixing, Osterman2011, chelakkot2021synchronized,cheng2024near,bennett2025direction}. While simplified frameworks often overlook the intricacies of filamentous motion, their utility in studying synchronization between individual cilia and the onset of metachronal waves is undeniable. Two such archetypes— the rower and rotor models— depict cilia or flagella as micron-scale beads, actively driven to follow either a circular/elliptical trajectory (the rotor) \cite{Andrej2006, Niedermayer2008, UchidaPrl10, liao2021energetics, mannan2020minimal, nasouri2016hydrodynamic, Brumley2016, Brumley2015, uchida2010synchronization, Meng2021, UchidaPrl2011, kotar2013optimal, Izumida2016} or a linear back-and-forth motion (the rower) \cite{CosentinoPre03, Bruot_prl2011, CosentinoSM09, CosentinoPRE12, Hamilton2021, StarkEpj11}. These models, admired for their simplicity and alignment with experimental observations \cite{CicutaPnas10, Bruot2016, Brumley2016, Maestro2018}, offer a foundational understanding. However, more realistic representations treat cilia as intricate filamentous structures, with dynamic regions responsible for the generation of active forces \cite{GueronPnas97, Guirao2007, Elgeti2013, Gilpin2020}. It is now well understood that hydrodynamic coupling between cilia can engender long-range metachronal waves, with the intricate interplay of activity, dissipation, and hydrodynamic forces central to wave formation and their properties \cite{Brato_pnas2022, brato2023collective, GolestanianPNAS2023, Deypre2018, Dey_pre2023}.

The axonemal beating has been explored before through the collective dynamics of coupled dynein motors, which stochastically attach to a filament with rates that are location-dependent and periodic along the filament~\cite{guerin2011dynamical, maggi2023thermodynamic, sharma2024active}. These molecular motors, once attached, actively extend, generating forces that sustain the oscillations of the filament. The upper bound to the beating precision is expected to be determined by the rate of energy dissipation, in accordance with the thermodynamic uncertainty relation \cite{barato_2015, Pietzonka2017, maggi2023thermodynamic, sharma2024active}.

We investigate how the interplay between cilia activity and dissipation due to the surrounding viscous medium influences precision of oscillation, a topic of recent interest \cite{maggi2023thermodynamic, sharma2024active}. To explore this, we first analyze the rower model numerically and then analytically in the symmetric case, quantifying precision using two related definitions of quality factor and linking it to fluctuations in the first passage time (FPT). We subsequently extend the analysis to encompass the asymmetric case. Our results show a clear data collapse and scaling of the quality factor, revealing a competition between activity and effective temperature. However, increasing noise strength causes deviations from the nice data collapse. Precision varies non-monotonically with oscillation amplitude, peaking at an optimal value. Applying a small noise approximation (SNA), we derive a closed-form expression for the FPT fluctuation and hence the quality factor, offering key insights into the role of activity and dissipation in precise beating. This expression explains the observed data collapse and scaling, and captures the precision peak at an optimal oscillation amplitude. As we show, the energy budget for achieving a desired precision, is consistent with the thermodynamic uncertainty relation.

\section{The Rower Model}

\begin{figure}[t!]
        \includegraphics[width=1.0\linewidth]{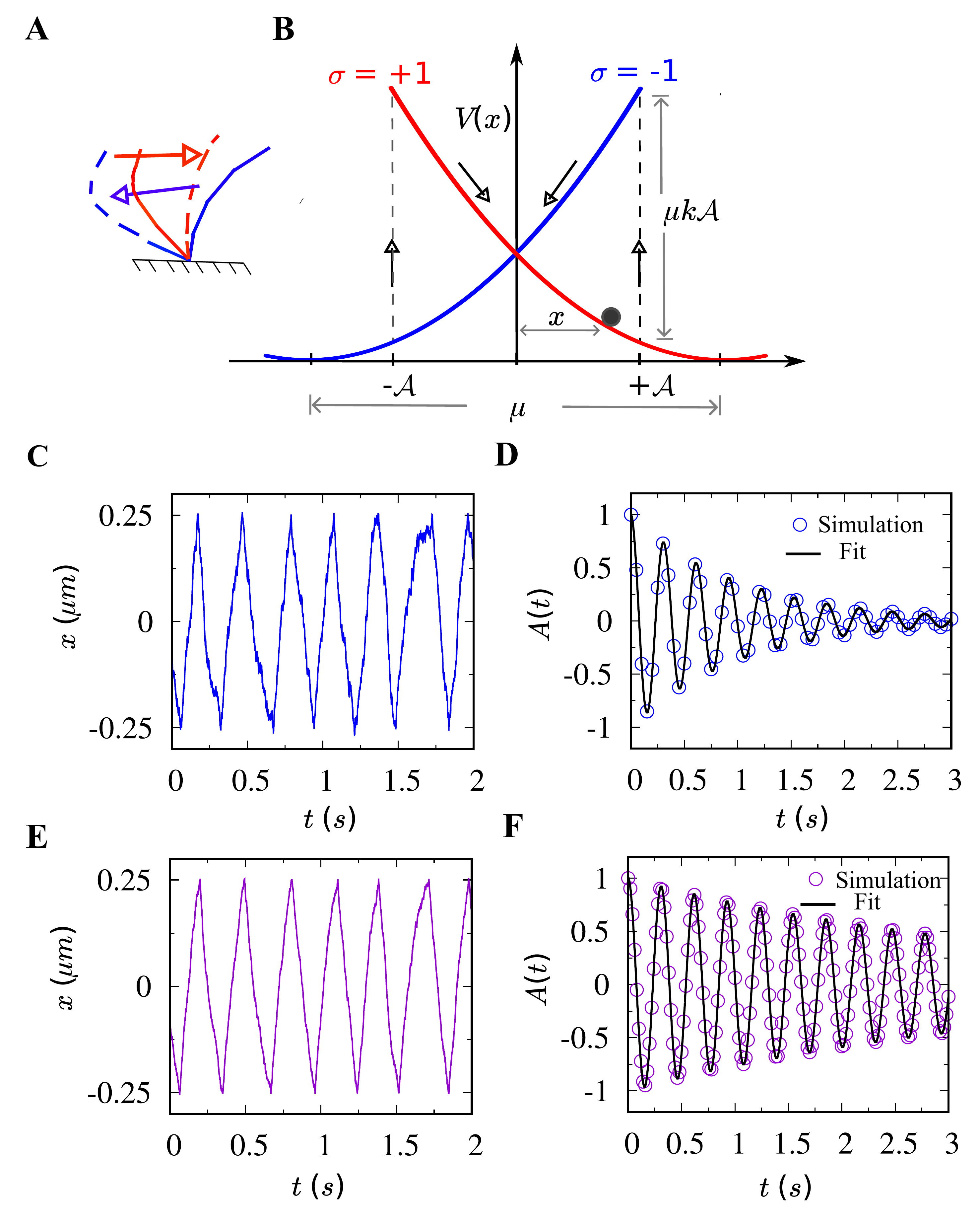}
\caption{Oscillations in the rower model for cilia. (A) A schematic of the realistic beating of a cilium is depicted. (B) It illustrates the oscillation mechanism in the rower model. A micron-sized bead in a fluid medium moves in the downhills of two harmonic potentials (represented by $\sigma=\pm1$) with stiffness $k$ and separated by a distance $\mu$. The motion is restricted between $-\mathcal{A}$ to $\mathcal{A}$ with $2\mathcal{A}<\mu$. Once the bead reaches a terminal position ($\pm\mathcal{A}$), switching between the potential happens, implying the pumping of energy $\mu k\mathcal{A}$. (C) and (E) Typical oscillating trajectories for high noise ($D=0.02\mu m^2/s$) and low noise ($D=0.005\mu m^2/s$)  strengths are shown. (D) and (F) Auto-correlation functions for trajectories for subplots (C) and (E) are shown. Other parameters used: $\mu=1 \mu m$, $\mathcal{A}=0.25 \mu m$, and $k=1.5pN/\mu m$. 
}
\label{Fig:1}
\end{figure}

In the rower model, the complex beating of a cilium is modeled as a one-dimensional periodic motion of a micron-sized bead in a fluid medium restricted within two specified locations (see Fig.~\ref{Fig:1}) \cite{CosentinoPre03, CicutaPnas10, Bruot_prl2011, CosentinoSM09, CosentinoPRE12, Bruot2016}. The oscillating motion in this over-damped system is generated by energy minimizing dynamics in two harmonic potentials and a position-based switching mechanism. The discrete variable $\sigma=\pm 1$ indicates the identity of the two potentials, with 
\bea
V(x,\sigma)&= \frac{kx}{2}(x-\sigma \mu).
\label{eq:potentials}
\eea
Here, $k$ is the stiffness, and $\mu$ is the separation between the minima of the two potentials, a natural length scale of the system. The motion is confined to the range $\pm \mathcal{A}$, where $2\mathcal{A} < \mu$. The bead undergoes Brownian motion only in the downhill regions of the potentials, and switching between the potentials at the terminal positions drives the bead in the opposite direction, creating sustained oscillations. Thermal fluctuations and other stochastic processes introduce noise into the oscillations. The over-damped motion of the bead is described by:
 \bea
        \frac{dx}{dt}  =  -\frac{1}{\gamma} \frac{\partial V(x,\sigma)}{\partial x}+\xi(t) \label{eqn:1}
                    \label{eq:dyn_rower}
\eea
where $\gamma=6\pi\eta a$ is the coefficient of drag force on the bead with radius $a$ due to the fluid with viscosity $\eta$, and $\xi(t)$ the white noise at time $t$. The noise satisfies $\la \xi(t) \ra=0$ and $\la \xi(t_1) \xi(t_2) \ra =2D\delta(t_1-t_2)$, where $D$ denotes a translational diffusivity. Within this model, the diffusivity and drag coefficient can be used to define an effective temperature  
\bea
k_{B}\Theta = D\gamma,  
\label{eqn:fd}
\eea
which, for active cilia, arises out of both thermal fluctuations and chemical processes involved in active energy pumping~\cite{Polin2009, Ma2014, Wan2014PRL, Quaranta2015}.
We investigate temporal coherence in oscillations of this system. 
When the bead reaches a terminal position ($\pm\mathcal{A}$), the potential switches, injecting energy $ \mu k\mathcal{A}$, which we refer to as the bead's activity.

Within a single potential well, the bead relaxes to equilibrium with a relaxation time $\tau_d = \gamma/k$, a natural time scale, in the absence of switching. The switching causes sustained oscillations of the cilium between $-\mathcal{A}$ and $\mathcal{A}$ in this overdamped dynamics. The average period $T_0$ of the oscillations is obtained by solving the mean dynamics for each $\sigma$ separately, and is given by:
\bea
T_0&=&2\,\tau_d\log\left(\frac{1+\mathcal{A}_s}{1-\mathcal{A}_s}\right),
\label{eqn:T0}
\eea
where $\mathcal{A}_s=2\mathcal{A}/\mu$ represents a dimensionless scaled amplitude. Note that the period increases linearly with viscosity $\eta$. Moreover, it increases with $\mathcal{A}_s$ to diverge at $\mathcal{A}_s=1$.



\section{Results}

In this section, we present definitions of quality factor describing the precision of oscillation and analyze the results, first numerically and then analytically within SNA.  

\subsection{Simulation details} 
We perform Euler-Maruyama integration of the stochastic differential equation Eq.\eqref{eqn:1} described above. 
We use integration time step equal to $10^{-4}$s, starting from random initial conditions. 
At $x=\mathcal{\pm A}$, the sigma value is flipped by taking $\sigma=-\sigma$. 
The simulation parameters in this paper are selected within the experimentally relevant range \cite{CicutaPnas10, Brumley2012}. We use $a = 1.5 \, \mu m$, and viscosity $\eta = 7.4 \, \text{mPa} \, \text{s}$, unless otherwise specified. The noise strength $D$ varies between $0.005$ and $0.50 \, \mu m^{2} \, \text{s}^{-1}$. In the presence of thermal noise alone, 
$D = k_B \Theta/ \gamma \approx 0.02 \, \mu m^2 \, \text{s}^{-1}$ at $\Theta = 300\,K$. Otherwise, $ k_B\Theta = D \gamma$ represents an effective temperature incorporating fluctuations due to active processes. Activity is varied with $k = 2 - 6 \, \text{pN} / \mu m$, $\mu = 1 - 3 \, \mu m$, and $\mathcal{A} = 0 - \mu / 2$. Example oscillating trajectories for high and low noise are shown in Fig.~\ref{Fig:1}(C) and (E), respectively.

\subsection{Precision quality factors of oscillations}
In deterministic systems, oscillations are perfectly coherent over time. Noise can disrupt this temporal coherence. The temporal precision, a dimensionless measure of oscillation coherence, is quantified by the quality factor $0 < Q < \infty$ \cite{remlein2022coherence, barato2017coherence, bagheri2014effects, Roldan_njp2021, placcais2009spontaneous, martin2001comparison}, which is defined as: 
\bea
Q=\frac{\tau_c}{T}.
\label{eq:Qdef}
\eea
For small fluctuations, the average period $T$ obtained from simulations closely matches the analytical formula in Eq.~\ref{eqn:T0}, as expected.

The auto-correlation function for displacement $x$, which quantifies the correlation of displacements at two different times, can be used to estimate $\tau_c$ and $T$. The normalized auto-correlation function $A$ is defined as:
\bea
    A(t)=\frac{\la x(t_0+t)x(t_0) \ra - \la x(t_0) \ra^2}{\la x(t_0)^2 \ra-\la x(t_0) \ra^2}.
\eea
The starting time $t_0$ should be large enough to avoid transients, and $\langle \cdot \rangle$ represents averaging over $t_0$ and trajectory ensembles. For noisy oscillations, $A(t)$ decays with lag time $t$ and can be fitted to $\cos(2\pi t/T) \exp(-t/\tau_c)$, allowing estimation of $\tau_c$ and $T$, and calculation of $Q$. A higher $Q$ indicates better precision. In Fig.~\ref{Fig:1}(C) and (E), typical trajectories for high ($D=0.02\ \mu m^2/s$) and low noise ($D=0.005\ \mu m^2/s$) are shown, with corresponding $A(t)$ in Fig.~\ref{Fig:1}(D) and (F). For high $D$, $Q \approx 3$, and for low $D$, $Q \approx 12$.

The quality factor $Q$ generally depends on the parameters $\mathcal{A}$, $\mu$, $k$, $\gamma$, and $D$. However, simple dimensional analysis shows that $Q$ depends only on the scaled amplitude $\mathcal{A}_s = 2\mathcal{A}/\mu$ and the dimensionless noise strength $1/\epsilon = D\gamma/(k\mu^2)$. Using $\tau_d = \gamma/k$ as the characteristic time and $\mu/2$ as the length scale, the rower dynamics (Eq.~\ref{eq:dyn_rower}) can be written in dimensionless form:
\bea
\notag
\dfrac{dx'}{dt'}=-(x'-\sigma)+\xi'(t')
\label{eqn:dyn_dimless}
\eea
where $x'=2x/\mu$ and $t'=t/\tau_d$ and the white noise satisfies $\langle \xi'(t_1') \xi'(t_2') \rangle = (8/\epsilon) \, \delta(t_1' - t_2')$. The potential switches at $x' = \pm \mathcal{A}_s$. While the average period $T$ depends only on $\mathcal{A}_s$ (Eq.~\ref{eqn:T0}), the coherence time depends on both $1/\epsilon$ and $\mathcal{A}_s$, so that $Q = Q(\mathcal{A}_s, \epsilon)$. Its explicit functional form, however, is nontrivial.

\begin{figure}[!]
        \includegraphics[width=1.0\linewidth]{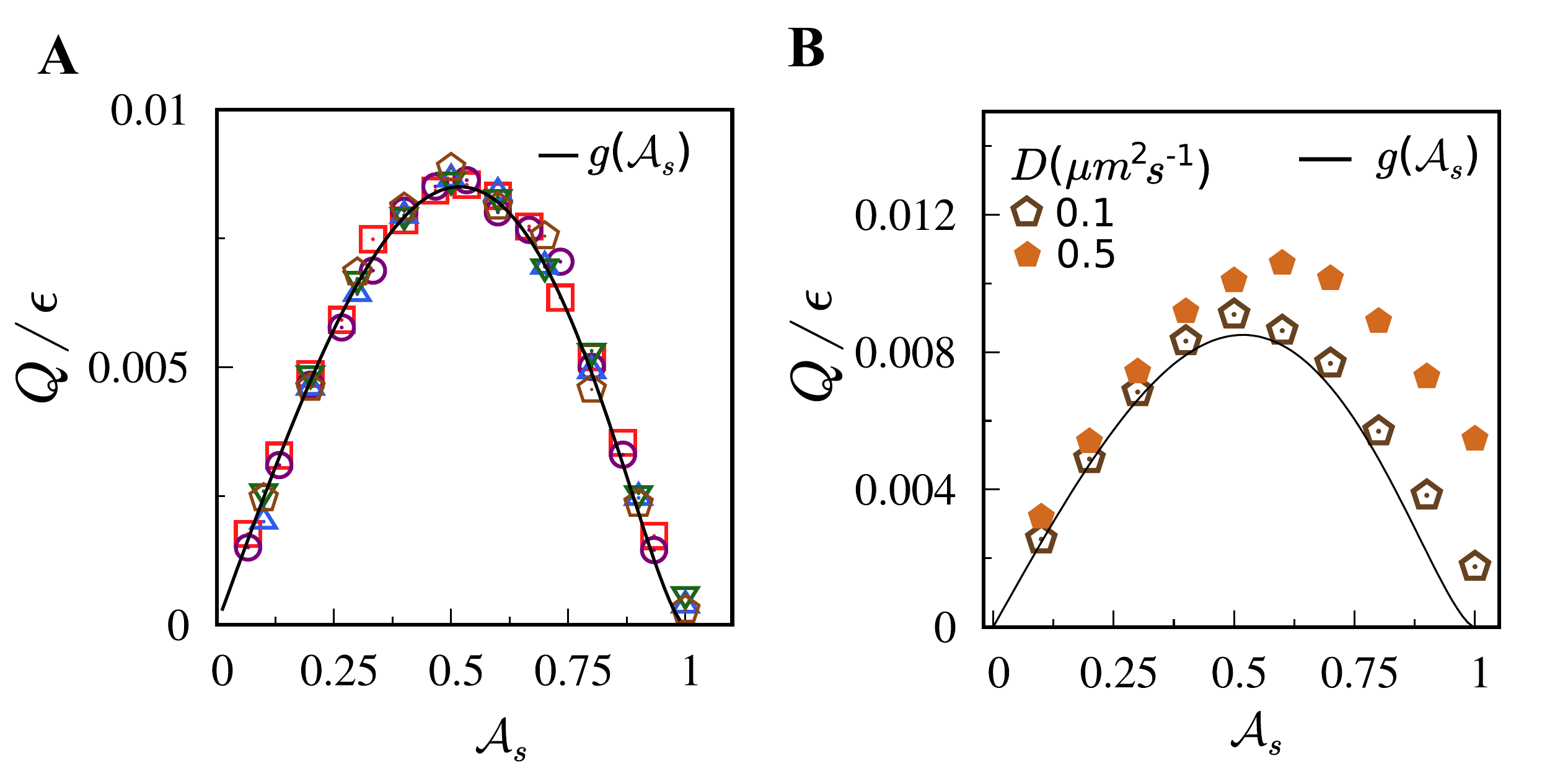}
\caption{The quality factor results for various activity and dissipation parameters are shown. (A) For small noise strengths: Scaled $Q$ vs  $\mathcal{A}_s$ for five different parameter sets collapse onto a single curve with the line representing the analytical formula in Eq.~\ref{eqn:g}. Parameters sets ($D$ in $\mu$m$^2$ s$^{-1}$, $\eta$ in mPas, $\mu$ in $\mu$m, and $k$ in pN/$\mu$m): (i) $0.02$, $7.4$, $3.0$, and $4.0$ (\scalebox{0.8}{$\bigcirc$}), (ii) $0.017$, $10.0$, $2.0$, and $6.0$ ($\triangle$), (iii) $0.017$, $10.0$, $2.0$, and $3.5$ ($\bigtriangledown$), (iv) $0.015$, $10.0$, $3.0$, and $2.0$ ($\square$), and (v) $0.01$, $7.4$, $2.0$, and $4$ (\scalebox{1.2}{$\pentagon$}). (B) For large noise strengths: Scaled $Q$ vs. $\mathcal{A}_s$ for large noise strengths, showing a deviation from data collapse. Parameters: $\eta{=}7.4$ mPas, $\mu{=}2\mu$m, and  $k{=}4$pN/$\mu$m.}
\label{Fig:2}
\end{figure}

Numerical results reveal two key features of the quality factor $Q$. First, $Q$ varies non-monotonically with $\mathcal{A}_s$, peaking at an optimal value (Fig.~\ref{Fig:2}). Second, for small noise, $Q$ data for different $\mu$, $k$, $\mathcal{A}$, $D$, and $\gamma$ collapse onto a single curve when scaled by $\epsilon$ and plotted against $\mathcal{A}_s$ (Fig.~\ref{Fig:2}(A)), indicating the scaling law
\bea
Q({\mathcal{A}_s, \epsilon})  = \epsilon g\left(\mathcal{A}_s\right).
\label{eq_Qscale}
\eea
The existence of such a scaling is not obvious. While the general small-noise ansatz is $Q({\cal A}_s, \epsilon) = \epsilon^\psi g({\cal A}_s / \epsilon^\chi)$, our data show the simpler form with $\psi = 1$ and $\chi = 0$. For large noise, this scaling breaks down (Fig.~\ref{Fig:2}(B)), with $g$ depending on both $\mathcal{A}_s$ and $\epsilon$. Thus, Eq.~\ref{eq_Qscale} holds only in the small-noise regime, relevant for axonemal beating.

\begin{figure}[t!]
 \centering
        \includegraphics[width=0.5\textwidth]{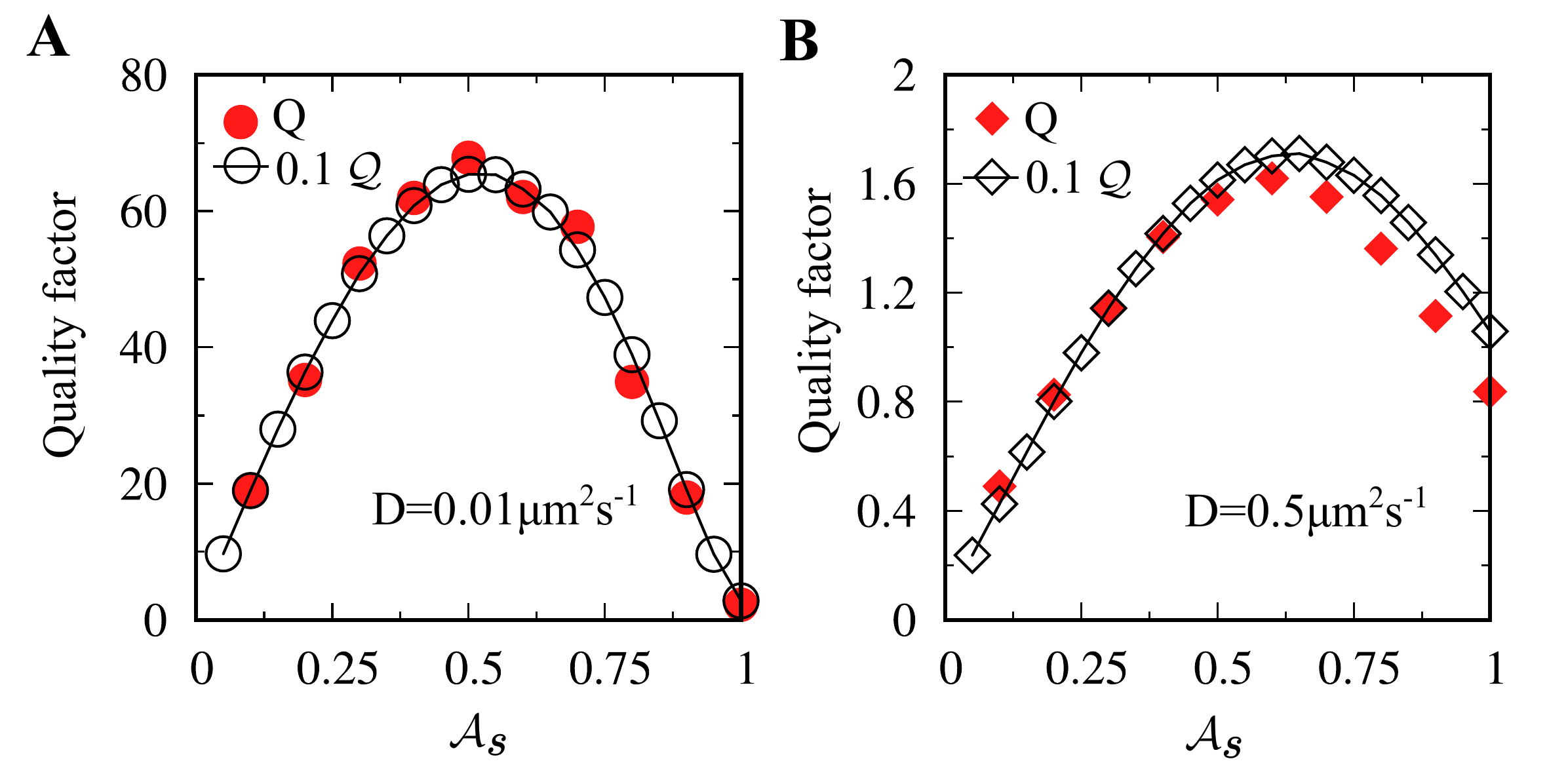}
\caption{The precision quality factors from two definitions are plotted against the scaled amplitude $\mathcal{A}_s$. Multiplying $\mathcal{Q}$ by 0.1 aligns the data, indicating the definitions differ only by a proportionality constant. Parameters: $\mu = 2 \, \mu m$ $k = 4 \, pN/\mu m$. The proportionality is more accurate at low noise (A) than at higher noise levels (B).}
\label{compare_Q}
\end{figure}

The interaction between activity and fluctuations plays a crucial role in determining the quality of the oscillations. Gaining analytical insight into the scaling behavior and the functional form of $g(x)$ is essential for a complete understanding of the system. To this end, we use a related measure of the quality factor~\cite{barato_2015, Pietzonka2017, maggi2023thermodynamic}:
\bea
{\cal Q} =  \frac{\langle J\rangle^2}{\langle J^2 \rangle - \langle J\rangle^2},
\label{eq:Q_CF}
\eea
where $J$ corresponds to a {\it current}, here, which is given by the oscillation velocity. 
The above quantity is used in recent literature to establish a thermodynamic uncertainty relation, which sets an upper limit based on dissipation.
The above definition of quality factor ${\cal Q}$ and $Q$ defined in Eq.~\ref{eq:Qdef} can differ by a proportionality constant, as shown in Fig.~\ref{compare_Q}.  
The position-dependent switching between potentials introduces non-linearity, making direct analytical calculations difficult. To simplify, we focus on the Brownian motion within each potential and define the current as $J = \frac{2\mathcal{A}_s}{\mathcal{T}}$, where $\mathcal{T}$ is the (dimensionless) time to travel between $-\mathcal{A}_s$ and $\mathcal{A}_s$ for $\sigma = 1$ (or vice versa for $\sigma = -1$). Within SNA, this leads to:
\bea
{\cal Q} = \frac{\langle \mathcal{T} \rangle^2}{\langle \mathcal{T}^2 \rangle - \langle \mathcal{T} \rangle^2} = \frac{1}{CV^2(\mathcal{T})}.
\label{eqn:Q_cv2}
\eea
The relative variance in first-passage time (FPT) is given by $CV^2(\mathcal{T}) = \frac{\langle \mathcal{T}^2 \rangle - \langle \mathcal{T} \rangle^2}{\langle \mathcal{T} \rangle^2}$~\cite{gardiner1985handbook, grebenkov2014first, redner2001guide, lindner2004moments, lindner2002maximizing}, offering a clearer understanding of the system's dynamics. FPT fluctuations, driven by switching events, introduce imprecision in the oscillations. By treating $\mathcal{T}$ as the first-passage time to reach $\mathcal{A}_s$ from $-\mathcal{A}_s$, we derive the formula for $CV^2(\mathcal{T})$ in the small fluctuation limit, as discussed in the following section. 

\subsection{First passage time (FPT) analysis}


Due to the symmetry of the potentials, the FPT statistics for one potential are identical to those of the other. Let $\mathcal{T}$ denote the FPT, the time (in dimensionless unit) for the bead to reach $\mathcal{A}_s$ for the first time from $-\mathcal{A}_s$ in the harmonic potential $k x^{\prime} (x^{\prime}-1)/2$. We aim to study the relative variance $CV^2$ in $\mathcal{T}$. The exact expressions of the relative variance $CV^2(\mathcal{T})$ is given by ~\cite{lindner2004moments,lindner2002maximizing}
\bea
\label{eqn:exact_cv2}
CV^2(\mathcal{T})&&=\frac{\la\mathcal{T}^2 \ra {-} \la  \mathcal{T} \ra^2}{\la \mathcal{T}\ra^2} \\\nonumber
&&=\frac{2\int_{y_-}^{\infty} {dy\, e^{y^2}[\operatorname{erfc(y)}]^{2}}\int_{y_-}^{y} {dz \,e^{z^2} \mathcal{H}(y_+{-}z)}}{\left[ \int_{y_-}^{y_+}  dy \; e^{y^2} \operatorname{erfc(y)} \right]^2 },
\eea
where $y_{\pm} = (1 \pm \mathcal{A}_s) / \sqrt{8/\epsilon}$, $\operatorname{erfc}(.)$ is the complementary error function, and $\mathcal{H}(.)$ is the Heaviside step function.
As this expression involves nested integrals with error functions, analytical insight is limited, and the dependence on $\mathcal{A}_s$ and $\epsilon$ is nontrivial.


If the fluctuations in $x^{\prime}$ are small and the dynamics of the mean and variance are known, then an approximated formula for uncertainty in FPT can be obtained using the latter. Using SNA, a method proven effective for FPT problems in gene expression \cite{kuheli_2024, co2017stochastic}, we derive the expression for the relative variance. Let us assume that the time $t'$ for the bead to reach a position $x'$ from an initial position fluctuates around the mean $\la t'\ra$, i.e., $t'=\la t' \ra + \delta t'$, where the fluctuation $\delta t'$ is small. Then, ignoring higher-order terms in the Taylor expansion of $x'(t')$, we get  
\bea
x'(t')&\approx& \la x'( t' )\ra + \delta t' \frac{dx'}{dt'}\Big|_{t'=\la t '\ra}.
\label{eqn:sna_taylor}
\eea
Taking the square of both sides of Eq.~\ref{eqn:sna_taylor} we obtain the expectation value
\bea
\la \delta x'^2\ra &\approx& \la \delta t'^2 \ra   \left(\frac{d\la x' \ra}{dt'}\right)^2\Big|_{t'=\la t' \ra},
\label{eqn:sna_exp}
\eea
where, $\delta x' = x'(t') - \la x'(t')\ra$. In terms of the squared coefficient of variation or relative variance, Eq.~\ref{eqn:sna_exp} can be written as
\bea 
\label{eqn:sna}
CV^2(\mathcal{T})\approx CV^2(x')\left(\frac{t'}{\la x'\ra }\frac{d\la x' \ra}{dt'}\right)^{-2}\Bigg|_{t'=\la \mathcal{T} \ra},
\eea
where $CV^2(x')=(\la x'^2 \ra - \la x' \ra^2)/\la x'\ra^2$, is the relative variance in displacement at the mean FPT.  Solving Eq.~\ref{eqn:1} for the first two moments in $x'(t')$, we get,  
 \bea
 \label{eqn:meanx}
&&\la x'(t') \ra {=} -\mathcal{A}_s \,\exp\left(-t'\right) {+} \left[1-\exp\left(-t'\right)\right],\\
\label{eqn:varx}
&&\la x'^2(t')\ra -\la x'(t')\ra^2 {=} \frac{4}{\epsilon}\left[1{-}\exp\left(-2t'\right)\right].
 \eea
Within mean-field approximation,  one obtains the mean FPT $\la \mathcal{T} \ra=T_0/(2\tau_d)$, half of the mean period. Using Eqs.~\ref{eqn:meanx} and ~\ref{eqn:varx} obtained at $t'=\la \mathcal{T}\ra$ in Eq.~\ref{eqn:sna}, we find  
\bea
\label{eqn:cvt}
 CV^2(\mathcal{T})=\dfrac{1}{\epsilon}\frac{16\,\mathcal{A}_{s}}{\left[(1-\mathcal{A}_{s}^2)\log\left(\frac{1+\mathcal{A}_{s}}{1-\mathcal{A}_{s}}\right)\right]^2}.
\eea                        
The expression above depicts a non-monotonic change in $CV^2(\mathcal{T})$ with $\mathcal{A}_s$, peaking at $\mathcal{A}_s = 0.518$.

 \begin{figure}[t!]
  \includegraphics[width=0.5\textwidth]{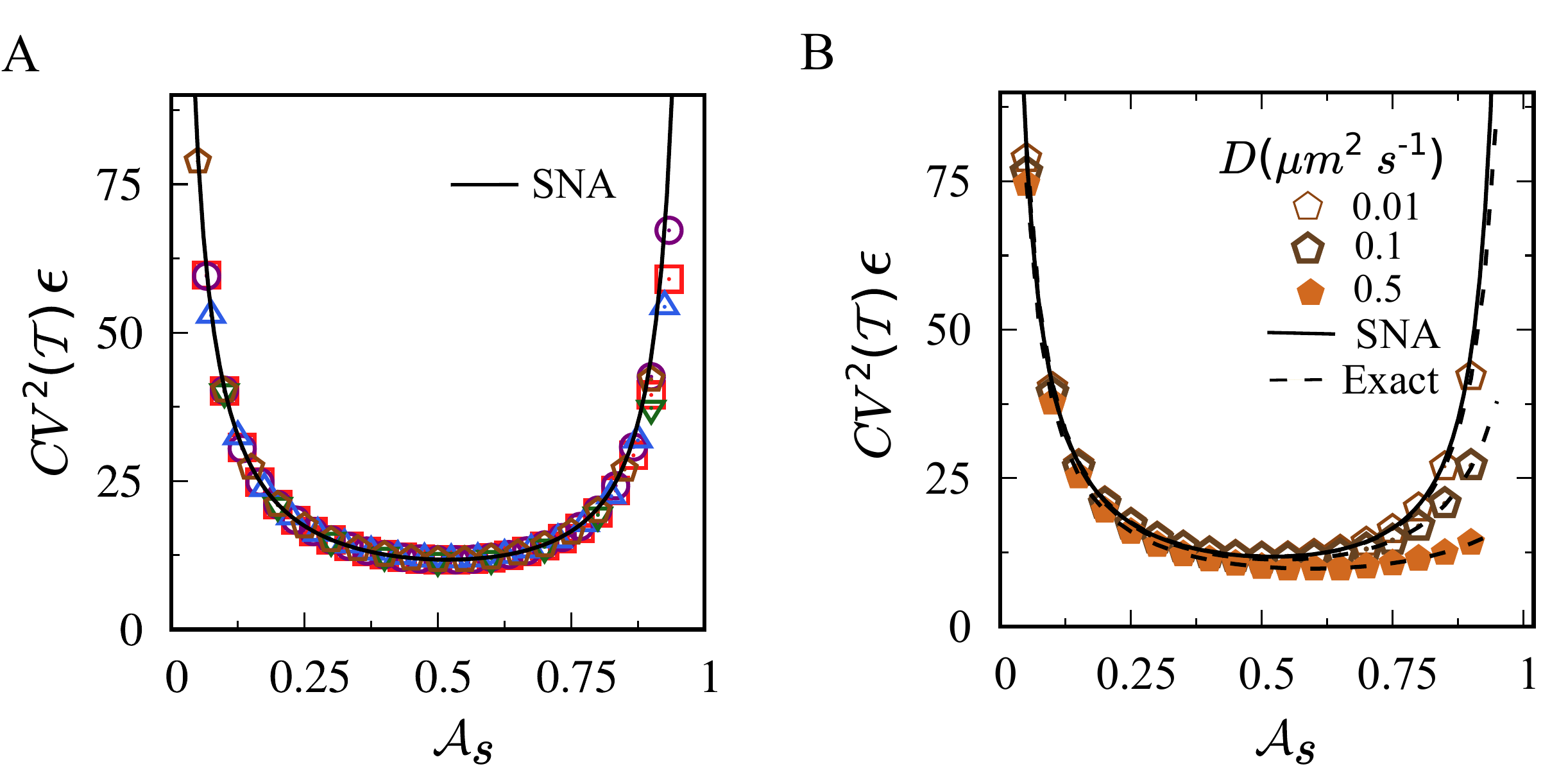}
\caption{Relative variance of FPT, $CV^2(\mathcal{T})$: (A) Scaled $CV^2(\mathcal{T})$ vs. $\mathcal{A}_s = 2\mathcal{A}/\mu$ for all parameters, with the black line showing the analytical formula (Eq.~\ref{eqn:g}). (B) Deviation from SNA results for higher noise strengths in the scaled plot. The dashed lines are obtained solving the exact formula given by Eq.~\ref{eqn:exact_cv2}.  Parameters as in Fig.~\ref{Fig:2}.
}
\label{Fig:3}
\end{figure}


For small noise strength, the simulation results of $CV^2({\cal T})$ for various values of $k$, $\mu$, $D$, and $\gamma$ collapse on to a master curve when scaled by the factor $\epsilon=\frac{k \mu^2}{D \gamma}$ and plotted against ${\cal A}_s$, as described by Eq.~\ref{eqn:cvt}, shown in Fig.~\ref{Fig:3}(A). Since $Q \propto 1/CV^2({\cal T})$, this data collapse for $CV^2({\cal T})$ also explains the collapse observed for $Q$ in Fig.~\ref{Fig:2}. The scaling function in Eq.~\eqref{eq_Qscale} is given by
\bea
\label{eqn:Q}
 g(\mathcal{A}_s) = \frac{\left[(1-\mathcal{A}_{s}^2)\log\left(\frac{1+\mathcal{A}_{s}}{1-\mathcal{A}_{s}}\right)\right]^2}{16\,\mathcal{A}_{s}},
 \label{eqn:g}
\eea   
up to a multiplicative constant. The plot of the scaling function in Fig.~\ref{Fig:2} (A) shows good agreement with numerical results. However, at higher fluctuations, a departure from the data collapse is observed in both $Q$ and $CV^2({\cal T})$, as shown in Fig.~\ref{Fig:2}(B) and Fig.~\ref{Fig:3}(B). In Fig.~\ref{Fig:3}(B), numerical results obtained from the exact expression, Eq.~\ref{eqn:exact_cv2}, show excellent agreement with Langevin simulations across both noise regimes. At large $D$, both results deviate simultaneously from the low-noise (SNA) prediction, indicating the breakdown of the SNA and the associated data collapse at high noise strengths.


\subsection{Energy dissipation rate}

Achieving high precision and the associated energy budget has been a key recent focus in biophysics research~\cite{Cao2015, Qian2007, Lang2014, Barato2016}. The thermodynamic uncertainty relation states the following inequality for the precision quality factor: ${\cal Q} \leq {\cal I}$ where ${\cal I} = \frac{q_{avg}}{2 D \gamma}$, where $q_{avg}$ is the dissipated heat over the driving time~\cite{barato_2015}, requiring a higher energy cost for a higher precision. However, as shown in ~\cite{Barato2016}, a cyclic external protocol driving a system to a periodic steady state can achieve high precision with an arbitrarily small energy budget.

\begin{figure}[t!]
 \centering
        \includegraphics[width=0.3\textwidth]{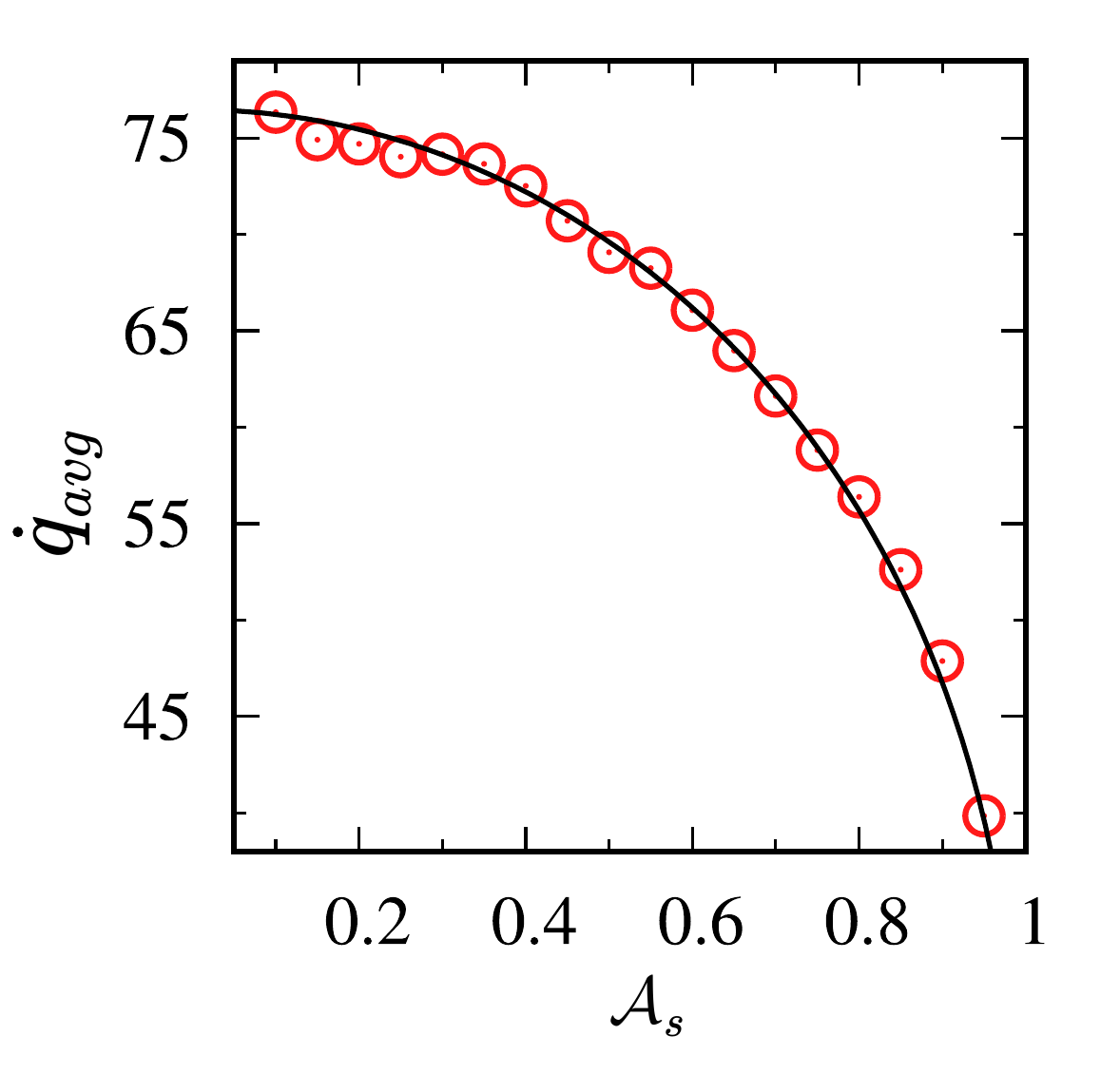}
\caption{Energy dissipation rate $\dot q_{avg}$ versus scaled amplitude $\mathcal{A}_{s}$. Symbols denote simulation results, and the line represents the analytical result (Eq.\eqref{eqn:qavg}). Parameters: $\mu = 2 \ \mu$m, $k = 4 \ pN/\mu$m, and $D=0.02\mu m^2 s^{-1}$. }
\label{Fig:4}
\end{figure}

Given this context, we ask about the energy budget required to achieve precision ${\cal Q}$ in the rower model. The energy input per cycle, through two switching events, is $2 k \mu \mathcal{A}$, which is fully dissipated. Due to symmetry, the dissipation rates for $\sigma=1$ and $\sigma=-1$ strokes are identical. The average energy dissipation rate during the movement from $-\mathcal{A}$ to $\mathcal{A}$ over the first passage time $\tau_d\mathcal{T}$ is given by:
\bea
\nonumber
\dot q_{avg} &=& \left\la \frac{1}{\tau_d {\cal T}} \int_0^{\tau_d\cal T}  {-}\frac{d V(x,\sigma=1)}{d x}\dot{x} dt\right\rangle \\ 
&\approx& \frac{k\mu\mathcal{A}}{\tau_d \la\mathcal{T}\ra}  = \frac{k^2 \mu^2 \mathcal{A}_{s}}{2\gamma \log\left(\frac{1+\mathcal{A}_{s}}{1-\mathcal{A}_{s}}\right)}.
        \label{eqn:qavg}
\eea 
In the final step, we use SNA. The average dissipation rate, $\dot q_{\text{avg}}$, increases with the activity parameters $k$ and $\mu$, but decreases monotonically with $\mathcal{A}_s$. For a comparison against numerical simulation results, see Fig.\ref{Fig:4}. For small ${\cal A}_s{\ll}1$, $q_{\text{avg}}$ behaves as:
\begin{align}
\dot q_{\text{avg}} &\approx \frac{k^2 \mu^2}{4 \gamma} \left(1-\frac{\mathcal{A}^2_s}{3}\right).
\end{align}
Despite energy dissipation rate decreasing monotonically with scaled amplitude ${\cal A}_s$, oscillation precision ${\cal Q}$ reaches its maximum at ${\cal A}_s = 0.518$, independent of other parameters.

As pointed out, the energy dissipated over an average period $\tau_d\la \cal T \ra$, $q_{avg} = \frac{1}{2} k \mu^2 {\cal A}_s$ and thus ${\cal I} 
= \frac{k \mu^2}{4 D \gamma} {\cal A}_s$, although ${\cal Q} = \frac{k \mu^2}{D \gamma} g({\cal A}_s)$. This leads to the relation 
\bea
\frac{{\cal Q}}{{\cal I}} = \frac{4 g({\cal A}_s)}{{\cal A}_s} = \frac{\left[(1-\mathcal{A}_{s}^2)\log\left(\frac{1+\mathcal{A}_{s}}{1-\mathcal{A}_{s}}\right)\right]^2}{4\,\mathcal{A}^2_{s}} .
\label{QbyA}
\eea
As the expression shows, $0 \leq \frac{{\cal Q}}{{\cal I}} \leq 1$, with the bounds defined by ${\cal A}_s = 1$ and ${\cal A}_s = 0$. This is consistent with the thermodynamic uncertainty relation, ${\cal Q}/{\cal I} \leq 1$. For any useful oscillation, ${\cal A}_s > 0$ is required, enforcing a stricter constraint, ${\cal Q}/{\cal I} < 1$. The exact expression for ${\cal Q}/{\cal I}$ in Eq.\eqref{QbyA} is obtained within SNA, but the upper bound to the ratio is more general.

\section{Precision in asymmetric beating of cilia}
Thus far, our analysis has assumed symmetric activity and dissipation in both beating directions. In reality, however, the forward and reverse beats of cilia are generally asymmetric. In this section, we examine how such asymmetry affects oscillation precision by studying an asymmetric rower model. The asymmetry may arise from differences in drag forces or from an asymmetric potential well \cite{CosentinoPre03, StarkEpj11}.

Here, we analyze the first-passage-time (FPT) statistics for the two branches separately. Let $\mathcal{T}_{\pm1}$ denote the dimensionless first-passage time for $\sigma = \pm1$. In the steady state, the total period is $\mathcal{T} = \mathcal{T}_{+1} + \mathcal{T}_{-1}$, and the fluctuations in the period can be written as,
\bea
CV^2(\mathcal{T}){=}\frac{CV^2(\mathcal{T}_{+1})\la \mathcal{T}_{+1}\ra^2{+}\, CV^2(\mathcal{T}_{-1})\la \mathcal{T}_{-1}\ra^2}{\la \mathcal{T}_{+1}{+}\mathcal{T}_{-1} \ra^2},
\eea
where $CV^2(\mathcal{T}_{\pm1})$ is the relative variance for $\sigma \pm1$.

{\it Asymmetric in potentials.}
Let the stiffness be $k+\alpha$ for forward beating ($\sigma=+1$) and $k-\alpha$ for reverse beating ($\sigma=-1$), with $\alpha<k$ quantifying the asymmetry ($\alpha=0$ corresponds to the symmetric case). The drag coefficient $\gamma$ is assumed to be identical for both branches. Using the SNA approach described above, the average period and its fluctuations are given by,
\begin{subequations}
\begin{align}
\la \mathcal{T} \ra_{\alpha}&=\dfrac{1}{\left(1-\frac{\alpha^2}{k^2}\right)}\la \mathcal{T} \ra_{0}, \text{ and } \\
CV_{\alpha}^2(\mathcal{T})&=\dfrac{\left(1+\frac{3\alpha^2}{k^2}\right)}{\left(1-\frac{\alpha^2}{k^2}\right)}CV_{0}^2(\mathcal{T}),
\end{align}
\end{subequations}
where $\la \mathcal{T} \ra_0$ and $CV_{0}^2(\mathcal{T})$, the mean time period and relative variance for $\alpha=0$ with
\bea
\nonumber
\la \mathcal{T} \ra_0{=}2\log\left(\frac{1{+}\mathcal{A}_s}{1{-}\mathcal{A}_s}\right), \text{ and } CV_{0}^2(\mathcal{T}) {=} \frac{1}{2\epsilon g(\mathcal{A}_s)}.
\eea
 Note that the asymmetry in stiffness enhances both the time period and the relative fluctuations about it.

{\it Asymmetric in drag forces.}
Here, we assume that the drag coefficients for forward and backward beating are $\gamma+\beta$ and $\gamma-\beta$, respectively, with $\beta<\gamma$ quantifying the asymmetry ($\beta=0$ corresponds to the symmetric case). The stiffness $k$ is taken to be the same for both beatings. The expressions for the average period and the associated noise then take the form:
\bea
\la \mathcal{T} \ra_{\beta}{=} \la \mathcal{T} \ra_{0}, \text{ and }
CV_{\alpha}^2(\mathcal{T}){=}\left(1{+}\dfrac{3\beta^2}{\gamma^2}\right)CV_{0}^2(\mathcal{T}).
\eea
As can be seen from the above expressions, asymmetry in the drag coefficient does not affect the average time period; however, it increases the relative fluctuations around it.

We note that the amplitude dependence of the fluctuation noise still remains governed by $g(\mathcal{A}_s)$, however,  asymmetry modifies the overall scaling through the parameters $\alpha$ and $\beta$.

\section{Comparison with the rotor model}

Finally, we briefly discuss the rotor model for cilia motion, where axonemal dynamics are approximated as circular motion of a colloidal bead under a tangential force $F$. The over-damped dynamics of the angular position $\theta$ of the rotor are given by
\bea
\frac{d\theta}{dt}  &=&  \frac{1}{\gamma R} F + \frac{1}{R}\xi(t) \label{eqn:dyn_rotor},
                    \label{eq:rotor_dyn_osc}
\eea
 where $\xi(t)$ is a Gaussian white noise with $\la \xi(t) \ra=0$ and $\la \xi(t) \xi(0) \ra = 2 D \delta(t)$.
 In rotor synchronization studies, the force $F$ must vary with $\theta$ \cite{uchida2010synchronization, UchidaPrl2011, Meng2021, Izumida2016, kotar2013optimal,Brumley2016, Maestro2018}, but for simplicity, here we assume $F = F_0$, a constant.
 The rotor model pumps energy continuously along the trajectory, unlike the rower model, which pumps at switching points. The rotor has a single length scale $R$, while the rower model involves two: $\mathcal{A}$ and $\mu$. The energy per cycle is $2\pi F_0 R$ for the rotor, and $2 k \mu \mathcal{A}$ for the rower.

Using SNA, the mean and relative variance for the time $\mathcal{T}$ to reach $\theta = 2\pi$ from $\theta = 0$ are given by
\bea
\la \mathcal{T}\ra_{\text {rotor}} = \frac{ 2 \pi R \gamma}{ F_0}, \text { and } CV_{\text{rotor}}^2(\mathcal{T}) = \frac{D\gamma}{\pi F_0 R}.
\label{eq:CV2_rotor}
\eea
The mean energy dissipation rate per cycle is
\bea
\dot q_{\text{avg, rotor}}&=& \frac{2 \pi R F_0}{\la \mathcal{T}\ra_{\text{rotor}}} = \frac{F^2_0}{\gamma}.
 \label{eqn:rotor_dissipation}
\eea
As $R$ increases, the oscillations become less noisy, but the average dissipation rate remains independent of $R$. However, both energy input and dissipation per cycle increase with $R$.

\section{Discussions and conclusion} 

Despite various uncertainties, cilia and flagella beat with high precision, an ability crucial for efficient microorganism locomotion and optimal fluid transport, such as in the respiratory tract. In this study, we used the rower model to examine oscillation precision, optimality, and the associated energy budget.
Given the nonlinear nature of oscillatory dynamics, analyzing precision is challenging. We addressed this by treating precision as a first-passage problem and applying SNA, which provided a simple formula for precision. The SNA is often valid for axonemal oscillations, shaped by their larger length and coherent activity.

The precision of oscillations is typically quantified by the quality factor $Q$, the ratio of coherence time to period. As we have shown, this is equivalent to another precision quality factor ${\cal Q}$, defined as the ratio of mean current to current fluctuation. This quantity is used in stochastic thermodynamics to establish a thermodynamic uncertainty relation, setting an upper limit based on dissipation. We use the SNA to express this quantity in terms of the relative variance in first passage time (FPT), $CV^2(\mathcal{T})$. In the rower model, the first passage time (FPT) $\mathcal{T}$ is the time for the beating to move from $-\mathcal{A}$ to $\mathcal{A}$ in a harmonic potential with stiffness $k$ centered at $\mu/2$ within a viscous medium with drag coefficient $\gamma$. The SNA provides a simple formula for $CV^2(\mathcal{T})$, matching numerical results closely. 
In the absence of a closed-form expression, the exact integral form of $CV^2(\mathcal{T})$ provides limited analytical clarity~\cite{Ahmad_pre2019, grebenkov2014first}, whereas the approximate SNA offers more profound insights.
We found that $CV^2(\mathcal{T})$ is a function of scaled amplitude $\mathcal{A}_s = 2\mathcal{A}/\mu$, with a scaling factor $k\mu^2/D\gamma$, representing the ratio of active energy scale to effective temperature. The SNA allows us to derive a closed-form expression of the scaling function $g({\cal A}_s)$ which reaches a maximum at ${\cal A}_s=0.518$, irrespective of other parameters. Our analytical expression for the ratio of precision quality factor ${\cal Q}$ and energy dissipation per cycle $q_{avg}$ gives a thermodynamic uncertainty relation ${\cal Q} \, (2D\gamma/q_{avg}) \leq 1$. 

Finally, we examined the effect of asymmetric beating on oscillation precision. We found that asymmetry lowers the overall precision relative to the symmetric case, while leaving the dependence on oscillation amplitude unchanged. The primary effect of asymmetry is to modify the overall scaling of the precision, which now depends explicitly on the asymmetry parameter.



Our predictions can be tested against experiments on cilia and flagella motion. The rower model predicts an optimization of precision quality with oscillation amplitude, which is absent in the rotor model. Testing our predictions can identify which model offers the simplest, meaningful description of actual dynamics. However, in real cilia or flagella, oscillation amplitude is determined by microscopic active processes, like motor protein dynamics, and their control may require involved biochemical techniques such as mutations or RNAi~\cite{Agrawal2003RNA, Alberts2002Molecular}. A more direct verification may be possible in colloidal systems mimicking axonemal beating~\cite{Bruot2016,Brumley2016,Maestro2018}.

\section*{Author Contributions} 
DC and SD designed and supervised the research. SG performed the analytical and numerical calculations. All authors participated in interpreting the results and contributed to writing the manuscript.\\


\section*{Data availability}
The data and codes are openly available at \url{https://github.com/guptasubhajit1995/Rower_precision.git}.

\section*{Acknowledgements}

DC acknowledges research grants from the Department of Atomic Energy (OM no. 1603/2/2020/IoP/R\&D-II/15028), an Associateship at ICTS-TIFR, Bangalore, and expresses gratitude to MPIPKS, Dresden for hospitality during a two-month visit in 2024, where part of the research was carried out. Simulations were carried out at computing facility HPCC Surya at SRM University -AP. SD thanks Estelle Pitard and Gladys Massiera for useful discussions. 





\bibliography{cilia_reference}{}

\end{document}